# The Prolog Interface to the Unstructured Information Management Architecture

Paul Fodor[1], Adam Lally [2], David Ferrucci [2]

[1] Stony Brook University, Stony Brook, NY 11794, USA,
pfodor@cs.sunysb.edu
[2] IBM T.J. Watson Research Center, Yorktown Heights, NY 10598,
{alally, ferrucci}@us.ibm.com

**Abstract.** In this paper we describe the design and implementation of the Prolog interface to the Unstructured Information Management Architecture (UIMA) and some of its applications in natural language processing. The UIMA Prolog interface translates unstructured data and the UIMA Common Analysis Structure (CAS) into a Prolog knowledge base, over which, the developers write rules and use resolution theorem proving to search and generate new annotations over the unstructured data. These rules can explore all the previous UIMA annotations (such as, the syntactic structure, parsing statistics) and external Prolog knowledge bases (such as, Prolog WordNet and Extended WordNet) to implement a variety of tasks for the natural language analysis. We also describe applications of this logic programming interface in question analysis (such as, focus detection, answer-type and other constraints detection), shallow parsing (such as, relations in the syntactic structure), and answer selection.

## 1 Introduction

The Unstructured Information Management Architecture (UIMA) is a framework for analyzing large amounts of unstructured content, such as text, audio and video [1-4], used for text and multi-modal analytics, question answering, bioinformatics, machine translation systems, knowledge integration, semantic search, etc. The innermost part of UIMA is the Common Analysis Structure (CAS), a dynamic data structure which contains: unstructured data (i.e., data whose intended meaning is still to be inferred), structured annotations over this data and various user views over these annotations. Applications based on the UIMA framework are composed of workflows of Analysis Engines, distributed software components with powerful search capabilities, which analyze the unstructured data and pre-annotations and assert new annotations in the CAS.

The UIMA Prolog interface is a generic Analysis Engine which translates the UIMA Common Analysis Structure (CAS) into a Prolog format, over which, the developers write rules and use top-down inference to search for new annotations over the unstructured data. In the following sections, we describe the encoding of the CAS into a Prolog Knowledge base and exemplify rules that exploit all the previous UIMA annotations (such as, the syntactic structure, semantic patterns) and external knowledge bases (such as, Prolog WordNet) to find new annotations in question analysis: focus detection, answer-type and other constraints detection, and shallow parsing. The QParse system is a question analysis application of the UIMA CAS Prolog interface. The QParse rules analyze questions in order to detect the focus (i.e., the segment of the question replaced by the answer) and the lexical answer-type (i.e., the minimum text span that identifies the semantic type of the correct answer, preserving singular vs. plural) using the syntactic structure (UIMA pre-annotations), lexical chains, the WordNet 3.0 external database [5]. An evaluation of QParse is compared with the previous version of the system (implemented without Prolog).

This document is structured as follows. The interface from the UIMA CAS to Prolog is introduced first, followed by the format of the data-structure and the retrieval of the results from Prolog as annotations in the UIMA CAS. Implementation details (such as: the API, some coding guidelines for the Prolog-based annotators and some data structure design decisions) are not included, but the used can easily access them on the UIMA framework Web site [2]. In the rest of the paper, we describe a set of use cases: shallow semantics detection of relations and the QParse component for question analysis operations (i.e., the focus and answer-type detection. We also evaluate these components' results.

## 2   The UIMA Prolog Generic Interface

The generic Prolog interface to UIMA is a generic analysis engine which can be specialized for particular tasks (such as, shallow parser, question analysis, and answer selection). This interface (see figure 1) translates the UIMA CAS into a knowledge base (see figure 2.a) dynamically asserted into the Prolog system, queries the unstructured data and the UIMA annotations found by analysis engines former to the current one in the workflow (see figure 2.b), and external Prolog knowledge bases (i.e., Prolog WordNet [5]) using a set of Horn clauses, and generates/inserts new annotations into the CAS.

Once an annotation is found and added to the annotations in the CAS, it can used to generate new annotations in the current analysis engine (e.g., in the shallow semantic parser) or analysis engines following the current one in the workflow. The transformed CAS is asserted dynamically for each specialized annotator and each data source, multiple Prolog annotators being concomitantly present in the workflow independent from each other. Currently, we have multiple Prolog analysis engines in our question answering system, for instance, shallow semantic parsing annotator (semantic relation finder), co-reference finder, question analysis annotator (focus and lexical aswer-type), etc. In the following section, we will exemplify a specific Prolog interface annotator for question analysis.

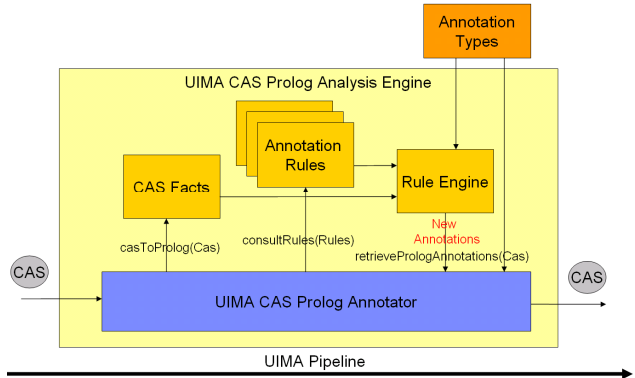

**Fig. 1.** The UIMA Prolog Analysis Engine Architecture

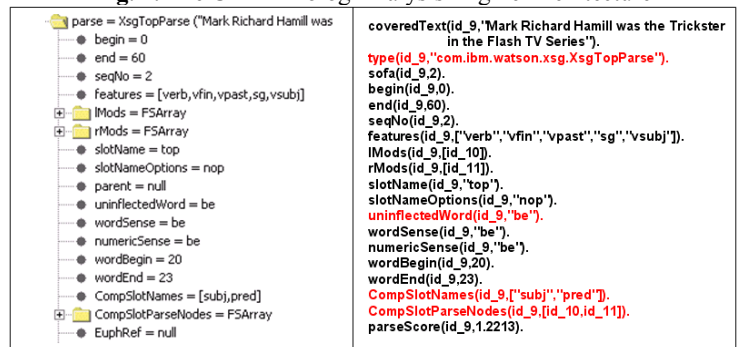

2.a. Translation of the UIMA CAS into Prolog facts dynamically asserted into the Prolog engine

```
castOf(Person,Composition):-
    lemmaForm(Verb,Text),
    wordNet:synonym(Text,["be","play","portray"]),
    subj(Verb, Person),
    semanticType(Person,"com.ibm.hutt.Person"),
    pred(Verb,Pred),
    pennTag(Pred,"NNP"),
    modifier(Verb,Mod),
    pennTag(Mod,"IN"),
    objprep(Mod,Composition),
    pennTag(Composition,"NN"),
    semanticType(Composition,"com.ibm.hutt.Composition").
```

2.b. A relation "*castOf*" detection rule for the shallow semantic parser

**Fig. 2.** Example of UIMA Prolog interface for shallow semantic parsing

## 3   The QParse Analysis Engine

The QParse analysis engine is a specialization of the generic UIMA Prolog interface used for natural language question-analysis. Its rules examine questions to detect the focus and the lexical answer type. The *focus segment* is the text span in a question that is replaced by the correct answer, while the *lexical answer-type* is the minimum text span that identifies the semantic type of the correct answer and preserves the number of the answer. It is used by many downstream analysis engines in a UIMA workflow for question-answering: proved based answer-selection, answer scoring components that identify the syntactic/semantic relations that must hold between the answer and other terms in the question, etc.

Qparse analysis engine has multiple inputs: the unstructured data input, the syntactic structure of the question (i.e., parsing pre-annotations), lexicalization pre-annotations (entity-name classifications), and uses external knowledge based (e.g., the Prolog WordNet 3.0). These inputs are transformed into Prolog facts and inserted into the Prolog engine (see figure 3). The Prolog engine applies a set of Horn clauses on this knowledge base and asserts new annotations in the UIMA architectures.

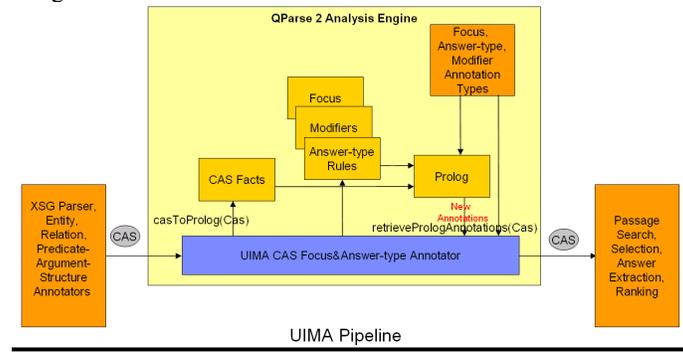

**Fig. 3.** The QParse analysis engine architecture in the question-answering system

QParse queries for annotations in the Prolog exploiting the syntactic structure in the questions. For instance, one pattern in the focus rule set searched for templates of "WHAT is the X …" (see figure 4.a), one pattern in the answer-type detection set, uses a lookup table of verb-objectType tuples (see figure 4.b second example).

| Pattern: WHAT IS X ...?<br>Example: "What is the democratic party symbol?"<br>    "What is the longest river in the world?"<br>Rule:<br>  focus(QuestionRoot, [Pred]):-<br>        getDescendantNodes(QuestionRoot,Verb),<br>        lemmaForm(Verb,"be"),<br>        subj(Verb,Subj),<br>        lemmaForm(Subj,SubjString),<br>        whatWord(SubjString), % e.g., "what","which" ("this","these")<br>        pred(Verb,Pred),!.<br><br>Pattern: HOW_MANY/MUCH X VERB ...?<br>Examples:  "How many hexagons are on a soccer ball?"<br>           "How much does the capitol dome weigh?"<br>Rule:<br>  focus(QuestionRoot, [Determiner]):-<br>        getDescendantNodes(QuestionRoot,Determiner),<br>        lemmaForm(Determiner,DeterminerString),<br>        howMuchMany(DeterminerString),!. % "how much/many" | Pattern: WHEN VERB OBJ; OBJ VERB THEN<br>Example: When was the US capitol built?<br>   answerType => ["types.Year"]<br>answerType(_QuestionRoot,FocusList,timeAnswerType,ATList):-<br>        member(Mod,FocusList),<br>        lemmaForm(Mod,ModString),<br>        wh_time(ModString), % "when", "then"<br>        whadv(Verb,Mod),<br>        lemmaForm(Verb,VerbString),<br>        timeTableLookup(VerbString,ATList),!.<br><br>Pattern: How ... VERB?<br>Example: "How did Virginia Woolf die?"<br>   answerType => ["types.Disease", "types.MannerOfKilling",…]<br>answerType(_QuestionRoot,FocusList,howVerb1,ATList):-<br>        member(Mod,FocusList),<br>        lemmaForm(Mod,"how"),<br>        whadv(Verb,Mod),<br>        lemmaForm(Verb,VerbString),<br>        howVerbTableLookup(VerbString,ATList), !. |
|---|---|
| 3.a. Examples of QParse focus detection rules | 3.b. Examples of QParse answer-type detection rules |

**Fig. 4.** Examples of QParse rules for question analysis

We evaluated the results of QParse on 412 manually annotated questions from the Text REtrieval Conference (TREC) and we got 370 correct matches (89.5%) and 17 super-types of the correct types (a better specialization was not found), while the rest of the annotations were wrong answers (either word sense disambiguation problems (for instance, we got the answer-type "*types.Facility*" instead of "*types.SportsTeam*" for the test "*What Liverpool club spawned the Beatles?*", the reason being that "*types.Facility*" meaning has higher probability than the "*types.SportsTeam*" meaning for the word "*club*" in the WordNet ontology) or parsing problems).

## 5   Conclusion

The UIMA generic Prolog annotator allowed us to develop faster and easier pattern matching rules for natural language analysis in a language familiar to our developers and users (i.e., the Prolog language), the Prolog engine being transparent to the UIMA pipeline (i.e., completely integrated in the pipeline), while having access to state-of-the-art semantics and proving effective on question analysis (i.e., time and results). We implemented interfaces for various rule systems: the UIMA-Sicstus Prolog interface (using the PrologBeans library) [6], the UIMA-SWI Prolog interface (using the JPL library) [7] and the UIMA-InterProlog translator for used by XSB [8] and Yap Prolog [9] systems (using the Interprolog library [10]). Our applications of this annotator include: complex rules for question analysis, shallow semantic parsing, and tools for development and testing UIMA analytics.